\documentclass[hidelinks,journal,twocolumn,superscriptaddress,floatfix,footinbib,notitlepage,groupaddress,showpacs]{IEEEtran}

\usepackage{multirow}
\usepackage{gensymb}
\usepackage{multicol}
\usepackage{graphicx}
\usepackage{amsmath}
\usepackage{amssymb}
\usepackage{graphics}

\usepackage{pdflscape}    
\usepackage{longtable}  

\usepackage{times,color,xcolor,amsfonts,amssymb,amsbsy,amsthm,amsmath,graphics,graphicx,hyperref,bm,bbm,makecell}

\usepackage{upgreek,hyperref}

\newcommand{\ignor}[1]{}

\newcommand{\ignore}[1]{}

\DeclareFontFamily{OT1}{pzc}{}
\DeclareFontShape{OT1}{pzc}{m}{it}%
{<-> s * [1.25] pzcmi7t}{}
\DeclareMathAlphabet{\mathpzc}{OT1}{pzc}%
{m}{it}

\let\oldsqrt\sqrt
\def\sqrt{\mathpalette\DHLhksqrt}
\def\DHLhksqrt#1#2{%
	\setbox0=\hbox{$#1\oldsqrt{#2\,}$}\dimen0=\ht0
	\advance\dimen0-0.2\ht0
	\setbox2=\hbox{\vrule height\ht0 depth -\dimen0}%
	{\box0\lower0.4pt\box2}}
\ifCLASSINFOpdf
\else
\fi

\begin{document}
	\title{Wireless Quantum Key Distribution in Indoor Environments }
	%
	%
	%
	
	\author{Osama~Elmabrok,~\IEEEmembership{Student~Member,~IEEE,}
		and Mohsen~Razavi 
		\thanks{This work was presented in part at the IEEE Globecom Conf. 2015 held in San Diego, CA. This research has partly been funded by the UK EPSRC Grants EP/M506953/1 and EP/M013472/1, and the ministry of higher education and scientific research in Libya. Both authors are with the School of Electronic and Electrical Engineering, University of Leeds, Leeds, LS2 9JT, UK (e-mail: elome@leeds.ac.uk and M.Razavi@leeds.ac.uk).}}
	\maketitle

\begin{abstract}
We propose and study the feasibility of wireless quantum key distribution (QKD) in indoor environments. Such systems are essential in providing wireless access to the developing quantum communications networks. We find a practical regime of operation, where, in the presence of external light sources and loss, secret keys can be exchanged. Our findings identify the trade-off between the acceptable amount of background light and the receiver field of view, where the latter specifies the type of equipment needed for the end user and its range of movements. In particular, we show that, using a proper setting, we can provide mobility for the QKD users without imposing stringent conditions on beam steering. 
\end{abstract}



%
\IEEEpeerreviewmaketitle

\section{Introduction}
%
%
%
%
\IEEEPARstart{Q}{uantum} key distribution (QKD) is a promising technology for achieving security in the quantum era~\cite{QKDReview2009}. It provides a secure way of distributing secret keys between two users over an optical channel~\cite{MDI400km, Zeilinger_07_Decoy}. The security is guaranteed by the properties of quantum mechanics rather than computational complexity~\cite{Gisin2002quantum}. The latter is at the core of public key cryptography schemes, whose security is threatened by advanced algorithms that can be run on quantum computers~\cite{shor1994algorithms,johnson2011quantum,barends2016digitized,moran2016quintuple}. QKD offers a solution to this problem and will possibly be the most imminent application of quantum technologies in our daily life. It is important to utilize the advantages of this scheme not only in niche markets but also for the general public~\cite{razavi2011multiple, razavi2012multiple}. This necessitates developing hybrid quantum-classical networks that support many users. This requires revisiting the requirements at both access and core parts of the network. This work focuses on the access networks, and, in particular, it addresses the wireless mode of access in indoor environments for such QKD networks. This would resemble a {\em quantum} Li-Fi system, possibly, in parallel with a classical Li-Fi, that enables end users to exchange secret keys with other network users in a convenient way \cite{GC2016}.

 The current dominant approach for ensuring data security over the Internet is based on a combination of public-key cryptography, e.g. the RSA protocol~\cite{rivest_1978_method}, for exchanging a secret key/seed, and symmetric-key cryptography protocols, such as advanced encryption standard (AES) or secure hash algorithm, for encryption, decryption, and authentication. The security of RSA is, however, based on the computational complexity of the factoring problem. The latter does not have any {\em known} efficient solutions on classical computers, but there exists a quantum algorithm using which one can solve the factoring problem in polynomial time~\cite{shor1994algorithms}. This is a huge threat to the security of our online communications, such as email correspondence and banking transactions, especially considering the progress made in the past few years in quantum  computing \cite{johnson2011quantum,barends2016digitized,moran2016quintuple}.  Note that, although symmetric-key algorithms such as AES may now be considered safe against quantum attacks~\cite{campagna2015quantum}, the initial input keys to such algorithms are currently distributed between two remote users using public-key schemes, which are vulnerable to quantum attacks. This would necessitate the implementation of alternative solutions, such as QKD, at large scale to offer data security to every individual user. 


Since its introduction by Bennett and Brassard in 1984~\cite{bennett1984quantum}, QKD has seen many field realizations \cite{SECOQC2009,TokyoQKDNetwork2011,Chinanetworks2009} with the capacity to generate verifiable security in data transmission even in the presence of an eavesdropper, Eve, who is assumed to have an unlimited computational power. In the original BB84 protocol~\cite{bennett1984quantum}, the light source was assumed to emit perfect single photons. However, in practice, it is now possible to use, without enduring a severe penalty, weak coherent laser pulses to approximate single-photon pulses. The progress with single-photon detectors has also been tremendous with quantum efficiencies as high as 93\% and dark counts as low as one per second are now achievable~\cite{marsili2013detecting}. Such developments have resulted in QKD being demonstrated over both optical fiber and free-space channels~\cite{MDI400km, Zeilinger_07_Decoy}. Today, various QKD systems are also commercially available. Examples are Clavis by ID Quantique in Switzerland~\cite{idquantique} and various products by QuantumCTek in China~\cite{QuantumCTek} that contributes to the 2000-km-long QKD link between Beijing and Shanghai. Other companies, such as Toshiba Research Europe Limited ~\cite{Toshiba_CRL} and British Telecom who are contributing to the UK national quantum network, are also heavily involved with the R\&D aspects of the technology. The latter examples represent QKD networks that are being developed across the world, in order to support a wider group of end users. The focus of most of these efforts is, however, mainly on the core networks~\cite{TokyoQKDNetwork2011}, or the wired access to such a backbone~\cite{frohlich2013quantum}. This work widens the QKD adoption by looking into \textit{wireless} indoor QKD.

Wireless access to a communications network is often taken for granted. This is not the case, however, for quantum communications. Most of QKD experiments are fiber-based in which point-to-point communication is established. In addition, the single photons enjoy traveling through a very low-loss channel. Through-the-air QKD experiments have also been point to point, therefore not offering mobility, and often require expensive and bulky optics equipment. However, nowadays, wireless connection is a necessity because of its convenience and also, because of the ever increasing use of handheld devices. In order that QKD will ever become ubiquitously used, {\em wireless mobile} QKD needs to be developed. While it is hard to envisage, at the moment, that wireless quantum access can be offered anywhere anytime, there are certain scenarios in which wireless QKD can be both possible and beneficial. For instance, customers in a bank office may wish to exchange secret keys with the bank wirelessly without the need for waiting for a teller or a cash machine. While conventional solutions based on direct optical links do not necessarily provide immunity against skimming frauds, QKD can guarantee the detection of any possible eavesdropping attacks \cite{HP_HandheldQKD}. Handheld prototypes have already been made, which enable a user to securely exchange a key with a cash machine \cite{HP_HandheldQKD,chun2017handheld}. It would be desirable to remove the constraint of being in the vicinity of a bank machine. In such a scenario, wireless indoor QKD is an attractive solution. In the long term, such indoors solutions can be part of a home/office network, which is equipped with wireless optical communications and is connected via fiber to its main service provider~\cite{GC2016}. This is in line with the developing Li-Fi technologies in data communications connected to passive optical networks for high data rate transmissions. Our proposed wireless QKD link will be using the same infrastructure while complementing the range of services offered to the user by adding quantum enabled security.

Wireless indoor QKD would not be without its own implementation challenges. Such a system is expected to suffer from the background noise, loss, and the implications of the mobility requirement. Ambient light, caused by the sun and artificial sources of light, is the primary hindrance to the successful operation of QKD in indoor environments. Essentially, QKD is a noise-dependent scheme, in which a secret key cannot be exchanged when the noise level exceeds a certain level. Another possible downside of indoor environments is the existence of severe loss, in comparison to fiber-based QKD, when the transmitted beam's angle is wide. For instance, the non-directed line-of-sight configuration, which suits most a mobile user, would suffer most from the background noise and loss. This is because, in this configuration, the receiver's field of view (FOV) should be sufficiently wide in order to collect sufficient power to operate. This would result in more background noise to sneak in. A wide beam angle at the transmitter would also result in a high channel loss, which may not be tolerable in the single-photon regime that most QKD schemes operate. Beam steering could be a possible solution to such problems, but it would add to the complexity of the system and the cost of handheld devices.

In this paper, we assess the feasibility of employing QKD in indoor environments by using the known techniques in optical wireless communications (OWC).  One distinctive feature of the setups we consider in our work is the use of phase encoding, which, as compared to polarization encoding used in \cite{HP_HandheldQKD,chun2017handheld}, offers higher resilience to alignment issues. This would, in principle, enables us to use wider beams and field of views, which are important factors for the mobility feature. Under such a setting, we find the feasibility regime for QKD in a single-room single-user scenario. Multiple users can also be supported by using relevant multiple-access techniques~\cite{razavi2012multiple}~\cite{bahrani2015orthogonal}. The system is mainly examined in the presence of background noise induced by an artificial lighting source, as well as the loss in indoor environments. We also account for possible imperfections in the encoder and decoder modules. Our results confirm that there exists a practical regime of operation where secret keys can be exchanged using wireless QKD links. Our results specify to what extent beam steering may or may not be needed. This investigation paves the road for any experimental realization of wireless indoor QKD~\cite{chun2017handheld}.

The remainder of this paper is organized as follows. In Sec.~\ref{Sec:Syst}, the system is described and in Sec.~\ref{Sec:Channel} the channel model is explained. We present secret key rate analysis in Sec.~\ref{Sec:Rate} and numerical results are discussed in Sec.~\ref{Sec:numeric}. Section~\ref{Sec:Conc} concludes the paper.

\section{System Description}
\label{Sec:Syst}
\begin{figure}[!t]
	\centering
	\includegraphics[width=.85\linewidth]{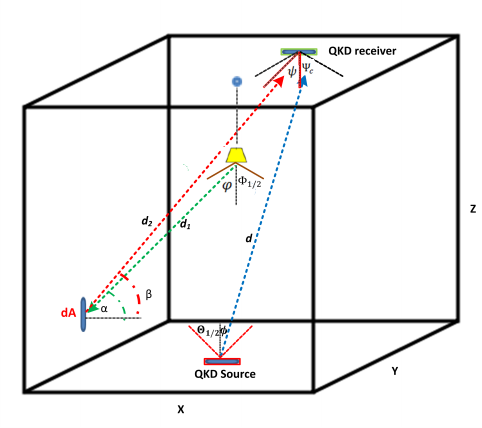}
	\caption{A wireless QKD link in an indoor setup. The transmitter is mobile, while the QKD receiver is fixed on the ceiling. For illustration purposes, the QKD receiver is depicted away from the center. In practice, it should be optimally placed at the center of the ceiling.}
	\label{fig_setup}
\end{figure}

In this section, we describe the setup and the components used in our wireless QKD system. Here we consider a particular scenario in which we have an empty window-less room of $X$$\times$$Y$$\times$$Z$ dimensions, which has been illuminated by an artificial source of light; see Fig.~\ref{fig_setup}. While this may not be exactly the case in a practical scenario, this particular setting allows us to properly study the resilience of the system to background noise. More realistic cases can also be investigated by properly adjusting the lighting source characteristics. The lighting source is assumed to be a Lambertian one with a semi-angle at half power of $\Phi_{1/2}$ located at the center of the ceiling. The contribution of the light source is calculated via its power spectral density (PSD), denoted by $S$, at the operating wavelength of the QKD link, denoted by $\lambda$. 

We assume that the QKD link is composed of two components. The QKD receiver, or Bob's box, is fixed and located at the center of the ceiling, while the QKD transmitter, or Alice's device, can be anywhere on the floor with a semi-angle at half power of  $\Theta_{1/2}$.  This setting has the practical advantage that the mobile user is only equipped with the encoder module for QKD, which is often less expensive and bulky than the decoder module. With regards to the QKD receiver and the artificial light source, we assume that the QKD receiver would just receive the reflected light from the walls and the floor and no light would enter the receiver directly from the bulb. This is achievable in practice by using certain reflectors that confine the radiation of the light source toward the floor. For simplicity, however, we assume that the lamp position is at the same level of the ceiling as the QKD receiver, and that the path between QKD transmitter and receiver is not blocked. We also implicitly assume that the QKD source shines light in an upward direction toward the ceiling. This requires a minimal alignment, which can be done by the users the same way that a mobile user may avoid being in a deep fading point when using their mobile phones. If the light beams used are not too narrow, then the total performance is expected to be tolerant of some movements. If they are narrow, however, then active beam steering would be required. We will see what range of beam angles we can use in our numerical results section. We assume that the QKD receiver has a detection area $A$ and an optical filter bandwidth $\Delta \lambda$. The unwanted light is filtered out best if the filter's bandwidth matches $1/T$, where $T$ is the width of the transmitted pulses by the QKD user.

We assume that the decoy-state variation of the BB84 protocol is in use~\cite{MXF:Practical:2005}. The key advantage of the decoy-state protocol is in allowing us to use weak laser pulses, instead of ideal single photon sources, at the source. This, while being immune to potential photon-number splitting (PNS) attacks~\cite{brassard_2000limitations}, offers a practical inexpensive solution for QKD encoders. 

{We employ time-bin encoding/decoding \cite{time-bin_encoding}, rather than polarization, to implement the BB84 protocol. In this scheme, the information is encoded onto the phase difference between two consecutive pulses. The possible advantage over polarization encoding is that we do not need to establish an identical polarization reference frame between a mobile device and the receiver. This can much simplify the alignment requirement and would allow us to use wider beams at the transmitter. We use passive decoders for our time-bin encoded signals, which includes a Mach-Zehnder interferometer with a delay corresponding to the time difference between the two time bins, followed by two single-photon detectors. The detectors are time gated so that they only pick up the signal when the two time bins are interfering.
	
	We consider three cases with regards to the QKD encoders. We first assume that the QKD encoding is carried out perfectly; that is, the phase difference between the two consecutive pulses is exactly as the protocol requires. We consider both cases of infinitely many and vacuum+weak \cite{MXF:Practical:2005} decoy states in this scenario. We also consider the case of using imperfect encoders. In this scenario, we assume that the device can be either characterized, in the sense that the QKD source has known flaws \cite{losstolerantarx}, or uncharacterized but has a fixed deviation from the ideal setting. In the latter case, we use the reference-frame independent (RFI) QKD protocol \cite{laing2010reference}, which has also been used in recent demonstrations of polarization-encoded handheld QKD \cite{HP_HandheldQKD,chun2017handheld}. The difference here is that, in the latter experiments, we need to at least know the reference for one polarization axis. In time-bin encoding protocols, the equivalent requirement would be to have two distinct time bins known to the receiver. The latter requirement is expected to be easier to achieve in practice for users that carry and move their QKD encoders. Other techniques have also been proposed to handle the basis mismatch problem, which may turn useful in certain scenarios \cite{mismatched_basis}. In section IV, the secret key rate analysis for each case will be given.

We consider the regime of operation when the reflected pulses off the walls are not overlapping in time with the main direct signal. This happens when the transmitted pulses are short in comparison to the transmission delay. This is the case in the practical regime of operation when high rate communications is desired. In this case, we neglect to collect the reflected QKD signals off the walls. We also assume that none of these reflected pulses will interfere with the forthcoming QKD pulses. That would imply that the repetition rate of the QKD link must be on the order of 100 MHz or lower, which is suitable for our scheme. 

Two scenarios are considered in this work. We first look at the case where the lighting source is turned off in which case, the background noise is assumed to be isotropic ambient light noise with a spectral irradiance denoted by $p_n$. The variation in the receiver's FOV in this case would affect the corresponding value of the channel transmittance of the QKD link. In the second scenario, we consider the effect of the artificial light source by accounting for the reflections from the walls and the floor, whose reflection coefficients are denoted by $r_1$ and $r_2$, respectively. The background noise at the QKD receiver, from the lighting source, will go up with increase in $S$ as well as that of the QKD receiver's FOV. The latter would determine how much mobility may be allowed. We therefore look at the trade-off between these two parameters in determining the secure versus insecure regimes of operations. Before doing that, in the following section, we first employ the propagation models in OWC for estimating path loss and background noise in the room.


\section{Channel Characterization}
\label{Sec:Channel}

Indoor environments can impose severe conditions for the operation of a QKD system, such as that in Fig. \ref{fig_setup}. This includes the issues of path loss, especially if a wide beam needs to be used, and the background noise, which also affects the performance of the scheme by increasing the error rate. In this section, we estimate the path loss and the background noise collected by the QKD receiver using established OWC models.

\subsection{Path Loss Estimation}

Due to path loss, the recipient, Bob, would receive a portion of the photons sent by the sender, Alice. This is estimated by the channel DC-gain, denoted by $H_{\rm DC}$, in OWC channels. The received power depends on the distance between the sender and the receiver, as well as the degree of directionality and the alignment. For the line-of-sight link between the QKD transmitter and receiver, the DC-gain is given by~\cite{kahn1997wireless}: 
\begin{equation}
\label{eq:HDC}
H_{\rm DC}= 
\begin{cases} 
\frac{A(m+1)}{2\pi d^2} \cos(\phi)^{m} T_s(\psi)  \\
\times  g(\psi) \cos(\psi),  ~~~~~~~~~~~~~~~~  0 \leq \psi \leq \Psi_c ,  \\ 
0 ~~~~~~~~~~~~~~~~~~~~~~~~~~~~~~~~~~ \text{elsewhere}, \end{cases}
\end{equation}
where $A$ is the detector physical area; $d$ is the distance between the QKD source and the QKD receiver; $\psi$ represents the incidence angle with reference to the receiver axis, while $\phi$ specifies the irradiance angle; see Fig. \ref{fig_setup}. These two parameters specify the relative location and orientation of the transmitter and receiver modules. $T_s(\psi)$ is the filter signal transmission; $m$ and $g(\psi)$ are, respectively, the Lambert's mode number used to specify the directivity of the source beam and the concentrator gain, given by
\begin{equation}
m=\frac{-\ln 2}{\ln(\cos (\Theta_{1/2}))}
\end{equation}
and 
\begin{equation}
g(\psi)= \begin{cases} 
\frac{n^2}{\sin^2(\Psi_{c})}, ~~~~~~ 0 \leq \psi \leq \Psi_c 
\\ 
0 ~~~~~~~~~~~~~~~~\psi > \Psi_c \end{cases},
\end{equation}
where $n$ is the refractive index of the concentrator, $\Psi_c$ is the receiver's FOV, and $\Theta_{1/2}$ is the semi-angle at half power of the light source. We neglect here the reflected pulses from the walls, which, as mentioned earlier, will arrive at a later time.

\subsection{Background Noise Analysis}

Two sources of background noise are accounted for in our analysis. Ambient light noise is considered first, which is due to black body radiation in the surrounding environment. The ambient light is assumed to be isotropic. The second source of background noise is the artificial lighting source in the room.

Let us first assume that the lighting source is off. The background noise in this case is due to isotropic ambient light. The received power for such isotropic ambient light is given by~\cite{kahn1997wireless}:
\begin{equation}
{P_\text{n,isotropic}}{(\lambda)}=p_n(\lambda)\Delta\lambda{T_s}{A}{n^2}.
\end{equation} 
The average number of detected photons over a time period $\tau$, corresponding to the width of a time bin, for each detector in a passive time-bin decoder is given by:
\begin{equation}
\label{nBiso}
n_{B}^{(1)}=\frac{{P_\text{n,isotropic}}{(\lambda)}\tau\eta_d /2}{hc/\lambda},
\end{equation}
where $c$ is the speed of light in the vacuum, $\eta_d$ is the single-photon detector efficiency, and $h$ is Planck's constant. Note that the background noise in two consecutive time bins, at the input of the decoder, would contribute to noise in the measured time bin at the output of the decoder. The passive decoder would, however, introduce an additional 1/2 loss factor, and the background noise is split evenly between the two detectors. The factor 1/2 in \eqref{nBiso} accounts for the aggregate effect.

Next, the background noise at the QKD receiver due to the lighting source is calculated. The light from the source can indirectly enter our QKD receiver via reflections off the walls and the floor. We obtain the amount of reflected power that may enter the QKD receiver by partitioning the floor and the walls into surface elements of area $dA$ and then calculating the contribution from each of these elements at the QKD receiver. The walls and the floor are modelled as diffuse reflectors, and we use the model presented in~\cite{gfeller1979wireless} to approximate the reflection pattern of the walls and the floor. Suppose the incident and reflected angles that the beams make with respect to the wall and the floor normal are $\alpha$ and $\beta$, respectively; see Fig. \ref{fig_setup}. The reflected beams would enter the QKD receiver if the receiving angle, $\psi$, with respect to the receiver normal, is less than $\Psi_{c}$. The portion of this reflected light that may enter the receiver is estimated by calculating the DC-gain for the reflected beam, $H_\text{Ref}$, and is given by \cite{gfeller1979wireless, ghassemlooy2012optical}:
\begin{equation}
H_\text{Ref}= \begin{cases} 
\frac{A(m_1+1)}{2\pi^2 d_1^2 d_2^2} \cos(\varphi)^{m_1}r T_s(\psi)g(\psi) \\
\times dA \cos(\alpha) \cos(\beta) \cos(\psi),   ~~~  0 \leq \psi \leq \Psi_c,   \\ 
0 ~~~~~~~~~~~~~~~~~~~~~~~~~~~~~~~~~~~ \text{elsewhere}, \end{cases}
\end{equation}
where $\varphi$ is the incidence angle with respect to the lighting source axis; $d_1$ and $d_2$ are, respectively, the distance from the lighting source to the surface element, and from there to the QKD receiver; $r$ is the reflection coefficient of the wall or the floor; and $m_1$ is the Lambert's mode number, which is calculated from (2), but with semi-angle at half power of $\Phi_{1/2}$ for the lighting source (rather than $\Theta_{1/2}$). By integrating over the walls and the floor, the average number of detected photons, per detector, at the QKD receiver due to the lighting source is then given by:
\\
\begin{equation}
n_{B}^{(2)}=\frac{S{(\lambda)}\Delta\lambda\tau\eta_d /2}{hc/\lambda}\int\int_\text{walls,floor}H_\text{Ref}.
\end{equation}

Different sources of background noise have different spectral irradiance over different ranges of wavelength. The ambient light caused by the Sun or an incandescent lamp covers a wide range of wavelengths and could generate a large number of background photons within a pulse period. In contrast, some common artificial sources, such as white LED bulbs, transmit mostly within the visible spectrum, possibly extended to the first window of infrared. For QKD systems operating at 880 nm of wavelength, the latter can then be a more tolerable source of noise. In order to accurately estimate the impact of white LED bulbs, we measured the irradiance of two randomly selected white LED bulbs, with an equivalent brightness to a 60-W incandescent lamp. The measurements were conducted by Photometric and Optical Testing Services. Spectral irradiance measurements have been done at a distance of 50 cm from the center of each bulb, from which the bulb's PSD has been calculated, see Fig.~\ref{fig_PSD_LED_bulbs}. The latter is measured to be on the order of $10^{-5}$ (W/nm) at 880 nm. {In this case, for the parameter values in Table~\ref{tablej}, the estimated value in (7), at FOV=20$^\circ$, is equal to $1.8 \times 10^{-5}$.} This is comparable to the dark count rate and it turns out, as will be shown later, that QKD operation can be feasible for such levels of external noise. The spectral irradiance for the sun is three orders of magnitude higher than that of the LED bulbs, hence the QKD system may only work under daylight exposure if the FOV is extremely narrow. On the other hand, the typical room temperature black-body radiation in the room is orders of magnitude weaker than that of LED bulbs.

\begin{figure}[!t]
	\centering
	\includegraphics[width=.95\linewidth]{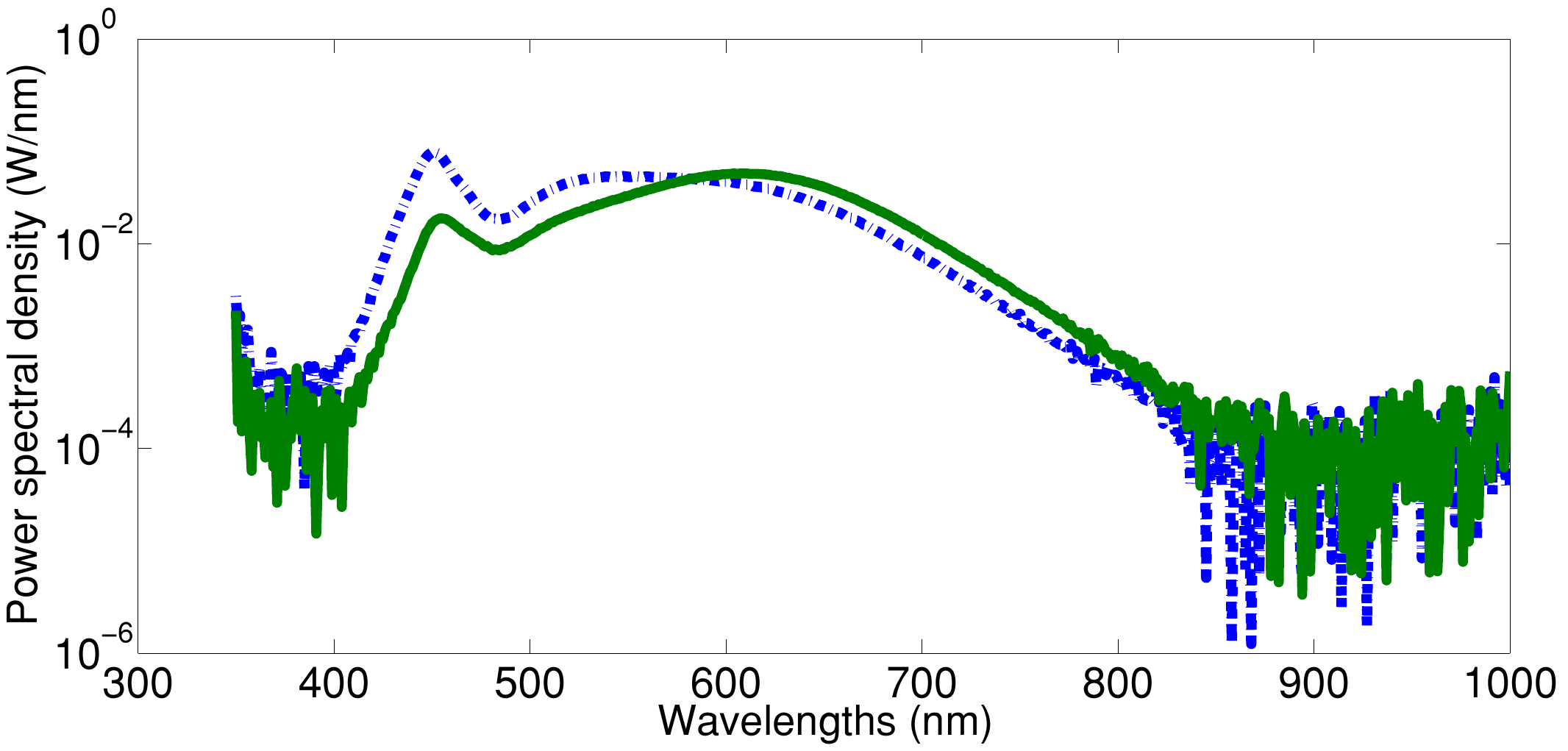}
	\caption{Power spectral density of two LED bulbs, equivalent to a 60-W incandescent lamp: A 650-lumen cool white LED manufactured by AURAGLOW (dashed), and a 805-lumen warm white LED bulb manufactured by INTEGRAL (solid). The wriggly form of the curves at the two far ends of the spectrum is due to the measurement precision.}
	\label{fig_PSD_LED_bulbs}
\end{figure}


\section{Secret Key Rate Analysis}
\label{Sec:Rate}
In this section, we present the rate analysis for our QKD system. The secret key generation rate, defined here as the probability of obtaining a secret key bit per transmitted quantum signal, is one of the key figures of merit for QKD systems. It will be affected by the amount of quantum bit error rate (QBER) of the system. The latter depends on the ambient noise and eavesdropping activities. QBER is defined as the ratio of non-identical bits in the sifted bits of Alice and Bob and the total number of sifted bits. In QKD protocols, if QBER is above a certain level, the protocol is aborted. In this section, we calculate the relevant key rate parameters for the three encoding techniques described in section \ref{Sec:Syst}, in the normal operating mode of the system, when there is no eavesdropper present.

\subsection{Decoy-state BB84 with perfect encoders}
\label{Sec:DecoyPerf}
The decoy-state technique, as mentioned earlier, was proposed in~\cite{hwang2003quantum} to combat the PNS attack~\cite{brassard_2000limitations}. The idea is that in addition to the signal pulses, Alice randomly sends decoy pulses with different intensities. This helps us gather more information about the quantum channel in order to discover the presence of Eve. Roughly speaking, as compared to the signal pulses, decoy pulses often have a lower multi-photon component, due to having a lower average number of photons per pulse as compared to the signal pulses. In this case, if Eve launches the PNS attack, Bob would receive a different portion of decoy pulses than signal pulses. As a result, if Alice and Bob examine separately both the decoy and signal pulses, the attack can be detected~\cite{MXF:Practical:2005}.

The secret key generation rate for the employed decoy-state protocol is lower bounded by~\cite{lo2005decoy,MXF:Practical:2005}  
\begin{equation}
\label{eq:R}
R_\text{decoy} \geqslant q\lbrace Q_1(1-h(e_1)-fQ_{\mu}h({E_\mu})\rbrace,
\end{equation} 
where $q$ is the basis sift factor, which is equal to $1/2$ in the original BB84 protocol. This is due to discarding half of the detection events in $X$ and $Z$ bases after the sifting procedure. Here, we use the efficient BB84 protocol~\cite{lo2005efficient}, which allows us to choose unevenly between $X$ and $Z$ bases, in which case $q$ can approach 1. If infinite number of decoy states are used, the parameters in \eqref{eq:R} are given as follows: $Q_1$, which is called the single-photon gain, is the probability that Alice sends one photon and Bob gets a click, and it is given by:
\begin{equation}
Q_1=Y_{1}\mu{e^{-\mu}},
\end{equation}
where $\mu$ is the average number of photons per pulse for the signal state and $Y_1$ is the yield of single photons. The latter is defined as the probability of getting a click provided that Alice has sent exactly a single photon, which is given by:
\begin{equation}
Y_1=1-(1-\eta)(1-n_N)^2,
\end{equation}
where $\eta$ is the total system transmittance given by $\eta_d H_{\rm DC}/2$ (the factor $1/2$ represents the loss incurred in a passive time-bin decoder), and $n_N$ is the total noise per detector given by $n_B+n_D$, where $n_D$ is the dark count rate per pulse for each of the two single-photon detectors at the Bob's receiver and $n_B=n_{B}^{(1)}+n_{B}^{(2)}$; $e_1$ is the error probability in the single-photon case, and is given by:
\begin{equation}
\label{eq:e1}
e_1=\frac{e_0{Y_1}-({e_0}-{e_d})\eta(1-n_N)}{Y_1},
\end{equation}
where $e_0=1/2$ and $e_d$ models the error, caused by channel distortions, in the relative phase between the two pulses generated by the phase encoder. In \eqref{eq:R}, $f$ represents the error correction inefficiency and $Q_\mu$ is the probability that Bob gets a click, when Alice sends a coherent state with an average number of photons $\mu$, and, it is given by \cite{panayi2014memory}:
\begin{equation}
Q_\mu=1-e^{-\eta\mu}(1-n_N){^2}.
\end{equation}
The overall QBER is represented by $E_\mu$ in \eqref{eq:R}, and it is given by
\begin{equation}
E_{\mu}=\frac{e_{0}{Q_\mu}-({e_0}-{e_d})(1-e^{-\eta\mu})(1-n_N)}{Q_\mu}.
\end{equation}
Finally, $h(x)$ is the Shannon binary entropy function given by
\begin{equation}
h(x)=-x\log_{2}x-(1-x)\log_{2}(1-x).
\end{equation}

If we use a two-decoy-state protocol, such as vacuum+weak, $Y_1$, $Q_1$, and $e_1$ are, respectively, bounded by~\cite{MXF:Practical:2005}   
\begin{align}
Y_\text{1} \geqslant Y_{1}^{L,\nu_1,\nu_2} & = \frac{\mu}{\mu \nu_1-\mu \nu_2-\nu_{1}^{2}+\nu_{2}^{2}}
[Q_{\nu_1}e^{\nu_1}-Q_{\nu_2}e^{\nu_2} \nonumber\\
& -\frac{\nu_{1}^{2}-\nu_{2}^{2}}{\mu^{2}}(Q_{\mu}e^{\mu}-Y_{0}^{L})],
\end{align}
\begin{align}
Q_\text{1} \geqslant Q_{1}^{L,\nu_1,\nu_2} & = \frac{\mu^{2}e^{-\mu}}{\mu \nu_1-\mu \nu_2-\nu_{1}^{2}+\nu_{2}^{2}}
[Q_{\nu_1}e^{\nu_1}-Q_{\nu_2}e^{\nu_2} \nonumber\\
& -\frac{\nu_{1}^{2}-\nu_{2}^{2}}{\mu^{2}}(Q_{\mu}e^{\mu}-Y_{0}^{L})],
\end{align}
and
\begin{align}
e_\text{1} \leq e_{1}^{U,\nu_1,\nu_2} & = \frac{E_{\nu_1}Q_{\nu_1}e^{\nu_1}-E_{\nu_2}Q_{\nu_2}e^{\nu_2}}{(\nu_1-\nu_2)Y_{1}^{L,\nu_1,\nu_2}},
\end{align}
where $\nu_1$ and $\nu_2$ are the average number of photons per pulse for the decoy states signals; $Q_{\nu_1}$ and $Q_{\nu_2}$ can be obtained from (12); $E_{\nu_1}$ and $E_{\nu_2}$ are the overall QBER for decoy-state signals given by (13). In the above equations, $Y_{0}$ is the probability of having a click due to the background and/or dark count noise, whose lower bound is given by
\begin{align}
Y_{0} \geqslant Y_{0}^{L} & =\text{max} \{\frac{\nu_1 Q_{\nu_2} e^{\nu_2}-\nu_2 Q_{\nu_1} e^{\nu_1} }{\nu_1-\nu_2},0 \}.
\end{align}
The above lower (L) or upper (U) bounds can be used in \eqref{eq:R} to find a lower bound on the key rate.

\subsection{Decoy-state QKD with known source flaws}

We investigate here the case of using non-ideal QKD encoders, for which the amount of deviation in state preparation is assumed to be known. This concerns a user/manufacturer that has characterized the QKD devices. In order to obtain a fair comparison with the previous case, we assume that the quantum states in the $Z$ basis, from which the secret keys are generated, are well aligned. In this basis, $|0 \rangle_Z$ and $|1 \rangle_Z$ represent single-photon states corresponding to the first and second time bin, respectively. For the sake of modeling the source flaws, however, we assume that the basis states in the $X$ basis, $|0 \rangle_X$ and $|1 \rangle_X$, are, respectively, given by $\cos(\frac{\pi}{4}+\frac{\delta}{2}) \vert 0\rangle_Z +\sin(\frac{\pi}{4}+\frac{\delta}{2}) \vert 1\rangle_Z $ and  $\cos(\frac{\pi}{4}+\frac{\delta}{2}) \vert 0\rangle_Z -\sin(\frac{\pi}{4}+\frac{\delta}{2}) \vert 1\rangle_Z $, where $\delta$ models the deviation from the ideal state. At $\delta =0$, $|0 \rangle_X$ and $|1 \rangle_X$ describe two consecutive pulses with phase differences of 0$^\circ$ and 180$^\circ$, respectively, as expected in the ideal case.

The asymptotic key rate in this case is given by~\cite{losstolerantarx}
\begin{equation}
R = Q_1[1-h(e_{x}^{(1)})]-Q_{\mu}fh(E_{\mu}),
\end{equation}
where $ Q_1$, $Q_{\mu}$, and $E_{\mu}$ are the same as those given for the decoy-state BB84. The phase error rate $e_{x}^{(1)}$ is expressed in terms of the conditional probabilities as~\cite{losstolerantarx}
\begin{align}
e_{x}^{(1)} & = \frac{Y_{1_\textit{X}|0_\textit{X}}+Y_{0_\textit{X}|1_\textit{X}}}{Y_{1_\textit{X}|0_\textit{X}}+Y_{0_\textit{X}|1_\textit{X}}+Y_{1_\textit{X}|1_\textit{X}}+Y_{0_\textit{X}|0_\textit{X}}}, 
\end{align} 
where, $Y_{s_{X}|j_{X}}$, for single-photon states, in our channel model is given by
\begin{align}
Y_{s_{X}|j_{X}} & = \eta [P_{s_{X}|j_{X}}(1-n_N)+\frac{1}{2}n_N]  \nonumber\\
& +(1-\eta)[n_N(1-n_N)+n_N^{2}/2], 
\end{align}
where $P_{0_{X}|0_{X}}=P_{1_{X}|1_{X}}=\frac{1}{2}[1+\sin(\frac{\pi}{2}+\delta)]$ and $P_{0_{X}|1_{X}}=P_{1_{X}|0_{X}}=\frac{1}{2}[1-\sin(\frac{\pi}{2}+\delta)]$.

\subsection{Decoy state QKD with unknown source flaws}

If the encoder imperfections are unknown, but correspond to fixed rotations, we can use the RFI-QKD protocol~\cite{laing2010reference}, instead of BB84, with the decoy-state technique. This case corresponds to scenarios where the users cannot characterize their devices and/or the manufacturers have not specified the extent of possible imperfections in the encoders. In the RFI-QKD protocol, one basis, $Z$, is supposed to be known and identical to both users, while $X$ and $Y$ bases can be different. In our time-bin encoding, the $Z$ basis is defined by each of the time-bin modes, whereas $X$ and $Y$ eigenstates are, respectively, given by superposition states $(|0\rangle_Z \pm |1\rangle_Z)/\sqrt{2}$ and $(|0\rangle_Z \pm i |1\rangle_Z)/\sqrt{2}$. We model the difference between the $X$ ($Y$) operator on Alice side, $X_A$ ($Y_A$), and that of Bob's side $X_B$ ($Y_B$) by a rotation parameter $\xi$, which gives us the following:
\begin{equation}
X_B=\cos(\xi) X_A+\sin(\xi) Y_A \ \ \text{and} \  \ Y_B=\cos(\xi) Y_A-\sin(\xi) X_A.
\end{equation}

In RFI-QKD, the correlation quantity, $C$, is used to estimate Eve's information, and it is defined by~\cite{laing2010reference}
\begin{equation}
C={\langle X_A X_B\rangle}^2+{\langle X_A Y_B\rangle}^2+{\langle Y_A X_B\rangle}^2+{\langle Y_A Y_B\rangle}^2,
\end{equation}
which can be written as~\cite{RFI_source_flaw}
\begin{equation}
C=(1-2E_{XX})^2+(1-2E_{XY})^2+(1-2E_{YX})^2+(1-2E_{YY})^2,
\end{equation}
where $E$ terms represent error rates in different scenarios. It can be shown that $C$ is independent of $\xi$. For our numerical analysis, $\xi$ is then assumed to be zero. In this case, we can assume that $E_{XY}$ = $E_{YX}$ = 1/2, and $E_{XX}$ = $E_{YY}$ = $E_{ZZ} =e_1$, where $e_1$ is the error probability in the single-photon case as calculated in \eqref{eq:e1}. The parameter $C$ can then be calculated by
\begin{equation}
\label{eq:C}
C=2(1-2e_{1})^2.
\end{equation}

Eve's information is bounded by~\cite{laing2010reference}
\begin{equation}
I_E=(1-e_1)h[\frac{1+\nu_{\max}}{2}]+e_1 h[\frac{1+f(\nu_{\max})}{2}], 
\end{equation}  
where $\nu_{\max}$ and $f(\nu_{\max})$, respectively, are given by
\begin{equation}
\nu_{\max}=\min[\frac{1}{1-e_1}\sqrt{C/2},1],
\end{equation}
and
\begin{equation}
f(\nu_{\max})=\frac{\sqrt{C/2-(1-e_1)^2 \nu^2_{\max}}}{e_1}. 
\end{equation} 

By substituting \eqref{eq:C} in the above equations, $I_E$ is then given by
\begin{equation}
I_E=e_1+(1-e_1)h[(1-3e_1/2)(1-e_1)].
\end{equation}

The key generation rate for RFI-QKD with decoy-state technique is then bounded by~\cite{RFI_source_flaw}
\begin{equation}
R \geqslant Q_{1}(1-I_E) - Q_{\mu} f h(E_{\mu}),
\end{equation}
where $Q_{\mu}$, $E_{\mu}$, $e_1$, and $Q_{1}$ are the same as those given for the decoy-state BB84 protocol.

\section{Numerical Results}
\label{Sec:numeric}
In this section, we numerically study the feasibility of wireless indoor QKD by looking at different scenarios. Table \ref{tablej} summarizes the nominal values used in our numerical results. These parameter values are based on the available technology for QKD systems. For instance, the values used for detector efficiency and dark count can be achieved by silicon APDs \cite{SPD_review}. Recent GHz-rate QKD demonstrations have also used pulse durations on the order of hundreds of ps with a correspondingly narrow filter at the receiver \cite{Toshiba_10G}. We use 0.5 photons per signal pulse, which is the near optimal value for $\mu$ in the decoy-state BB84 protocol. The room dimensions are representing a typical room with higher reflections from the walls than the floor, which might have carpeting. In a large partitioned office space, this can represent one cubicle in the room, or the area that can be covered by one QKD receiver. We also assume a rather large semi-angle at half power for both the lighting source and the QKD source, which overestimates the errors we may have in a practical setup. The latter will, however, be crucial for mobility features.

\begin{table} 
	\caption{Nominal values used for our system parameters. }        
	\label{tablej}
	\center
	\begin{tabular}{ccc}
		\hline\hline    
		Symbol	 &  Parameter&  Values   \\
		\hline 		
		$\Phi_{1/2}$ {} & Semi-angle at half power of the bulb&  $70\degree$ \\
		$\Theta_{1/2}$& Semi-angle at half power of QKD source&  $30\degree$ \\
		$\lambda$		& Wavelength of QKD source& $880$~nm \\
		$X , Y , Z$& Room size&4,4,3 m    \\
		$r_1,r_2$		& Reflection coefficients of the walls and floor &  $0.7$ , $0.1$ \\
		$\eta_{d} $ &Detector efficiency& $0.6$    \\
		$\tau$ & Pulse width& $100$~ps    \\
		$\Delta\lambda$ & Optical filter bandwidth& $\frac{\lambda^{2}}{\tau c}$    \\
		$A$ & Detector area& $1$ cm$^2$     \\
		$n$		& Refractive index of the concentrator&  1.5 \\
		$T_{s}$ & Optical filter transmission& 1   \\
		$n_{D}$ & Dark count& $1000\tau$   \\
		$\mu$ & Average no. of photons per signal pulse& $0.5$   \\
		$\nu_1 , \nu_2 $ & Ave. no. of photons/pulse for decoy states& $0.1 , 0$   \\
		$f$ & Inefficiency of error correction& $1.16$   \\
		$e_d$ & relative-phase distortion probability& $0$   \\
		\hline
	\end{tabular}
\end{table}

Before analyzing the key rate performance, let us start by illustrating how the position of the QKD source and its beam size, as well as the receiver's FOV, would affect the path loss. Figure~\ref{fig_total_loss} shows that the position of the QKD source with respect to its receiver can harshly affect the channel loss. In our case, moving from the center of the room to one of its corners adds nearly 10~dB to the channel loss. This is partly because of the additional distance, but mainly because of the loose alignment we have adopted for our system. In our setup, we have assumed that the QKD source emits light upward, at a right angle with the floor, to the ceiling. There is also the interplay between the source semi-angle and the total loss. The lower the source semi-angle is the larger $m$ would be in \eqref{eq:HDC}. This means that for small values of $\phi$, i.e., when the QKD source is near the center of the room, there may be some gain in the channel transmissivity proportional to $m+1$, but once $\phi$ increases, by moving toward the corners, that benefit may be washed away by the $\cos(\phi)^m$ term in \eqref{eq:HDC}. The receiver's FOV would also affect the loss in a negative way. One may assume that a larger FOV would result in higher power collection, but, quite the opposite, because of the concentrator gain, the larger the FOV is, the higher the channel loss would be. This is evident from all the curves in Fig.~\ref{fig_total_loss}. A larger FOV would, however, enable the QKD source to be seen over a wider range. That advantage could, however, come with additional background noise that may sneak into the receiver. Overall, Fig.~\ref{fig_total_loss} indicates that, without beam steering, the QKD system may need to tolerate a large amount of propagation loss in the wireless channel.

\begin{figure}[!t]
	\centering
	\includegraphics[width=.95\linewidth]{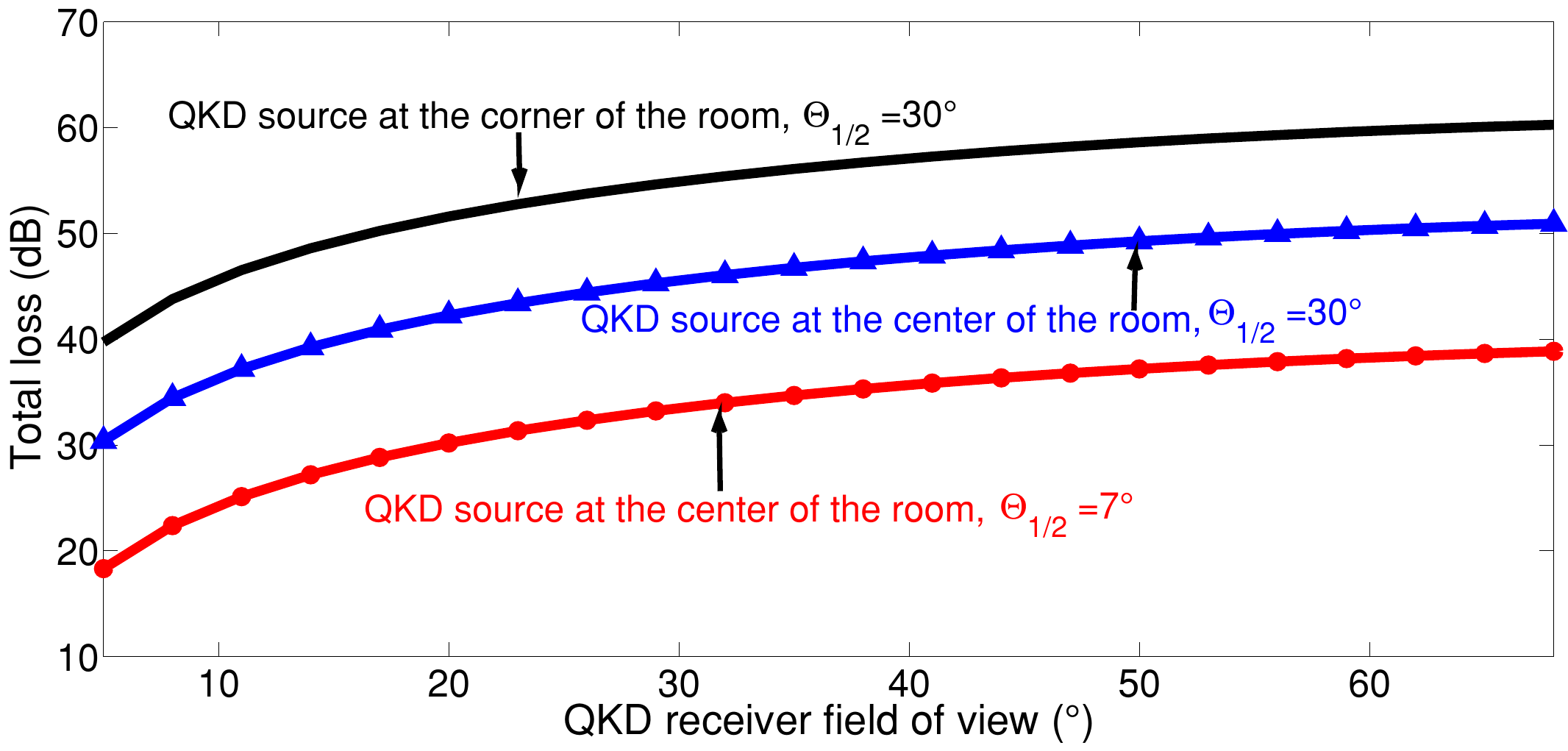}
	\caption{Total loss, $-10\log_{10}(\eta_d H_{\rm DC}/2)$, for the QKD source in center and corner positions. Path loss depends on the semi-angle at half power, the position of the QKD source, and the receiver's FOV.}
	\label{fig_total_loss}
\end{figure}

In the following, we first assess the in-principle feasibility of wireless QKD using the decoy-state BB84 protocol with perfect encoders. We then consider the effect of encoder imperfections in our analysis.

\subsection{Secure versus Insecure Regions}

In this section, two scenarios of wireless QKD in indoor environments are examined, and the corresponding secret key generation rates are obtained using loss and background noise calculations in the previous sections. In the first scenario, only the background noise due to the isotropic ambient light is included. The key generation rate is then presented versus a range of FOVs as the latter has an impact on path loss as shown in Fig.~\ref{fig_total_loss}. In the second scenario, we account for the background noise induced by the lighting source. In such a case, the amount of background noise would depend on the PSD of the lighting source and the QKD receiver's FOV. In both cases, we use the decoy-state encoding with infinitely many decoy states as described in Sec.~\ref{Sec:DecoyPerf}.


\begin{figure}[!t]
	\centering
	\includegraphics[width=.98\linewidth]{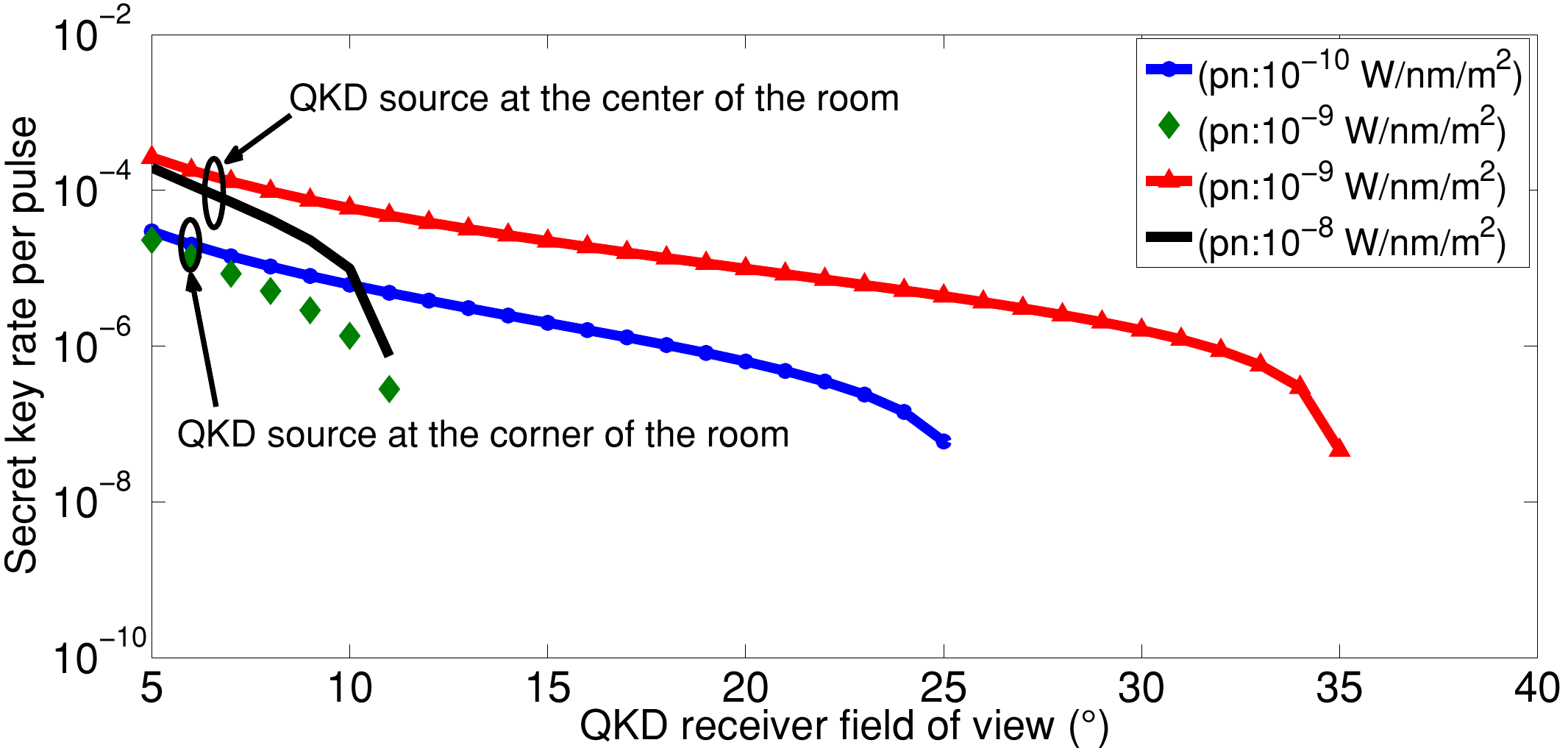}
	\caption{Secret key rate per transmitted pulse when the lighting source is off, and the background noise is only due to the ambient noise. The QKD source is sending light upward with $\Theta_{1/2}=30\degree$. {The decoy-state BB84 protocol with an infinite number of decoy states and perfect encoders are employed here.}}
	\label{fig_key_rate_ambient_noise}
\end{figure}

Figure \ref{fig_key_rate_ambient_noise} shows the secret key generation rate in a dark room when the QKD transmitter is either at the center of the room, or at a corner. In both cases, the QKD receiver is fixed at the center of the ceiling. When the QKD source is located at the center of the room, the system would tolerate spectral irradiance of ambient noise on the order of $10^{-8}$~W/nm/m$^2$, as shown in Fig.~\ref{fig_key_rate_ambient_noise}. Due to additional loss, the scheme would tolerate less ambient noise when the QKD source is located at a corner of the room. The ambient light noise considered in this case is due to black body radiation in the surrounding environment, whose effect is calculated via \eqref{nBiso}. This implies that the resulting background noise is constant in this case, and the drop in rate in Fig.~\ref{fig_key_rate_ambient_noise} is merely because of the loss in the system as seen in Fig.~\ref{fig_total_loss}. The spectral intensity emitted at room temperature (300 K) from objects, such as the human body, is expected to be low at the operating wavelength of our scheme. According to Planck's formula, the spectral irradiance at room temperature at 880 nm is on the order of $10^{-18}$~W/nm/m$^2$, which is far below the tolerable amount of ambient noise in our scheme. This is mainly because we assume that there would be no sun light in the room, which can adversely affect system performance. In the case of our window-less room, however, it is safe to neglect the effect of isotropic ambient noise in our system. We next consider the effect of the lighting source on our QKD operation.

\begin{figure}[!t]
	\centering
	\includegraphics[width=.95\linewidth]{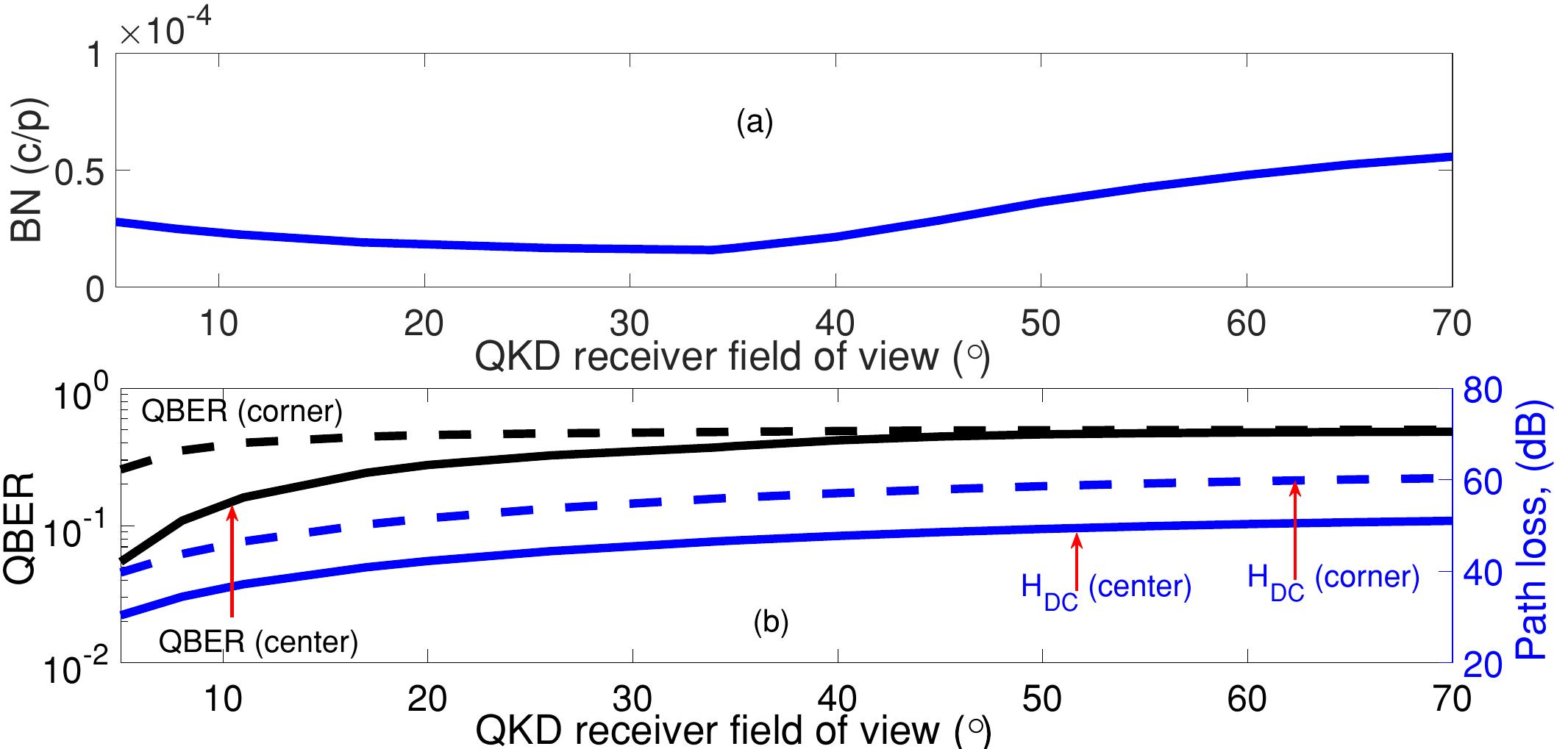}
	\caption{(a) Background noise (BN) in count per pulse (c/p), generated by the artificial light source, collected by the QKD receiver versus FOV. (b) The channel loss, $1/H_\text{DC}$, and QBER, $E_{\mu}$, versus FOV. The QKD source is located either at the center or corner of the room floor. The PSD of the bulb is $10^{-5}$~W/nm. }
	\label{fig_fov_versus_Hdc_QBER}
\end{figure}

When the lighting source is on, additional background noise would sneak into the QKD receiver. The choice of FOV to some extent affects the amount of background noise. Figure~\ref{fig_fov_versus_Hdc_QBER} shows the effect of FOV on the background noise, as well as on $H_{\rm DC}$ and the overall QBER, $E_\mu$. It is interesting to see that, at the beginning, the background noise would slightly drop until FOV reaches near $34\degree$, from which point it gradually increases with FOV. For narrow FOVs, the concentrator gain is rather high, but it sharply approaches 1 when FOV increases. This can be seen in the behavior of $H_{\rm DC}$ as well. The drop in the concentrator gain can justify the initial decline of the background noise. Another contributing factor is that for FOV $<34\degree$, the collected power at the QKD receiver is mainly induced by the reflection from the floor, whereas for FOV $>34\degree$, the reflection from the four walls would also matter. We have chosen much higher reflection coefficient from the walls (0.7) than the floor (0.1), which justifies the increase in the background noise. From the QBER curves in Fig.~\ref{fig_fov_versus_Hdc_QBER} it can be seen that the QBER is larger than its acceptable threshold for large FOVs. It is then fair to assume that within the region of interest for the FOV, the background noise is nearly constant while the channel gain drops with the increase in the FOV.

Figure \ref{fig_key_rate_3D_center} shows the secret key generation rate per transmitted pulse when the lighting source is on and the QKD source is placed at the center of the room's floor facing up toward the receiver. The figure shows the trade-off between the receiver's FOV and the PSD of the light source. The higher the PSD is, the lower the FOV should be in order to improve the signal to noise ratio at the QKD receiver. This restriction partitions the $x$-$y$ plane in Fig.~\ref{fig_key_rate_3D_center} into secure and insecure regions. The insecure region is when the lower bound on the key rate is zero, i.e, when the secure exchange of keys cannot be guaranteed. The secure region then specifies when QKD operation is feasible within the setting in our setup. Figure~\ref{fig_key_rate_3D_center} implies that with a PSD of $10^{-5}$ W/nm, corresponding to white LED bulbs, the QKD receiver's FOV should be less than 7$^\circ$. 

\begin{figure}[!t]
	\centering
	\includegraphics[width=1\linewidth]{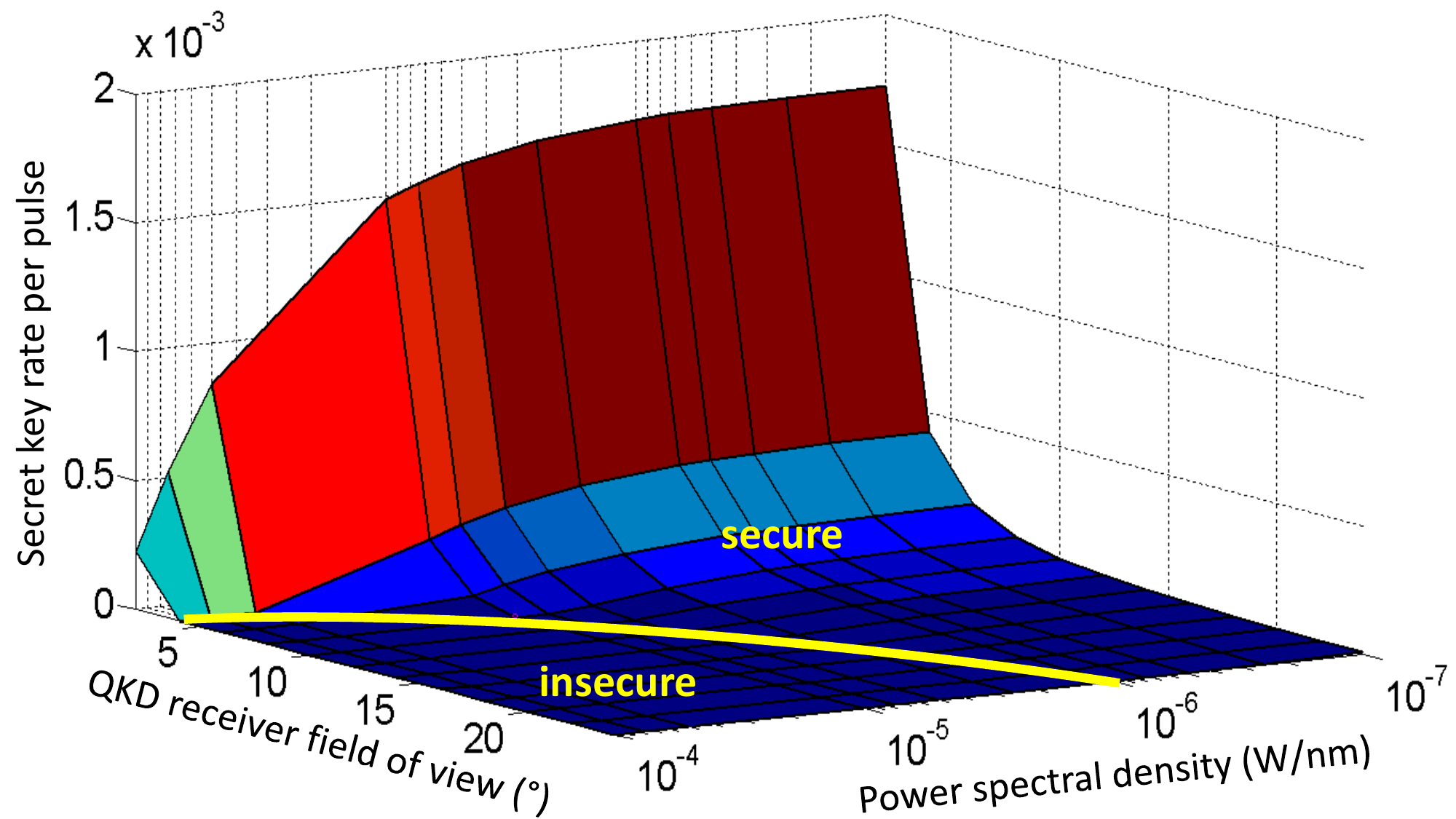}
	\caption{Secret key rate per transmitted pulse for a QKD source ($\Theta_{1/2}=30\degree$) at the center of the floor in the presence of a lighting source. {The decoy-state BB84 protocol with an infinite number of decoy states and perfect encoders are assumed here.}}
	\label{fig_key_rate_3D_center}
\end{figure}

Figure \ref{fig_key_rate_3D_corner} shows the secret key generation rate per transmitted pulse when the transmitter has moved to a corner of the room. We assume that the QKD source has a rather large transmission angle (30$^\circ$), in which case a portion of its beam has the chance to be received by the receiver. This can possibly be achieved by a diffuser, if the source beam is too narrow. We assume that the source is again sending light up toward the ceiling. Being at the corner of the room, the channel DC-gain is lower than that of a transmitter at the center of the room. This would imply that lower amounts of background noise can be tolerated in this case. The trade-off between the FOV and PSD has been shown in Fig.~\ref{fig_key_rate_3D_corner}. In this case, at a PSD of $10^{-5}$ W/nm, the QKD receiver's FOV should be less than {4$^\circ$}. 

\begin{figure}[!t]
	\centering
	\includegraphics[width=1\linewidth]{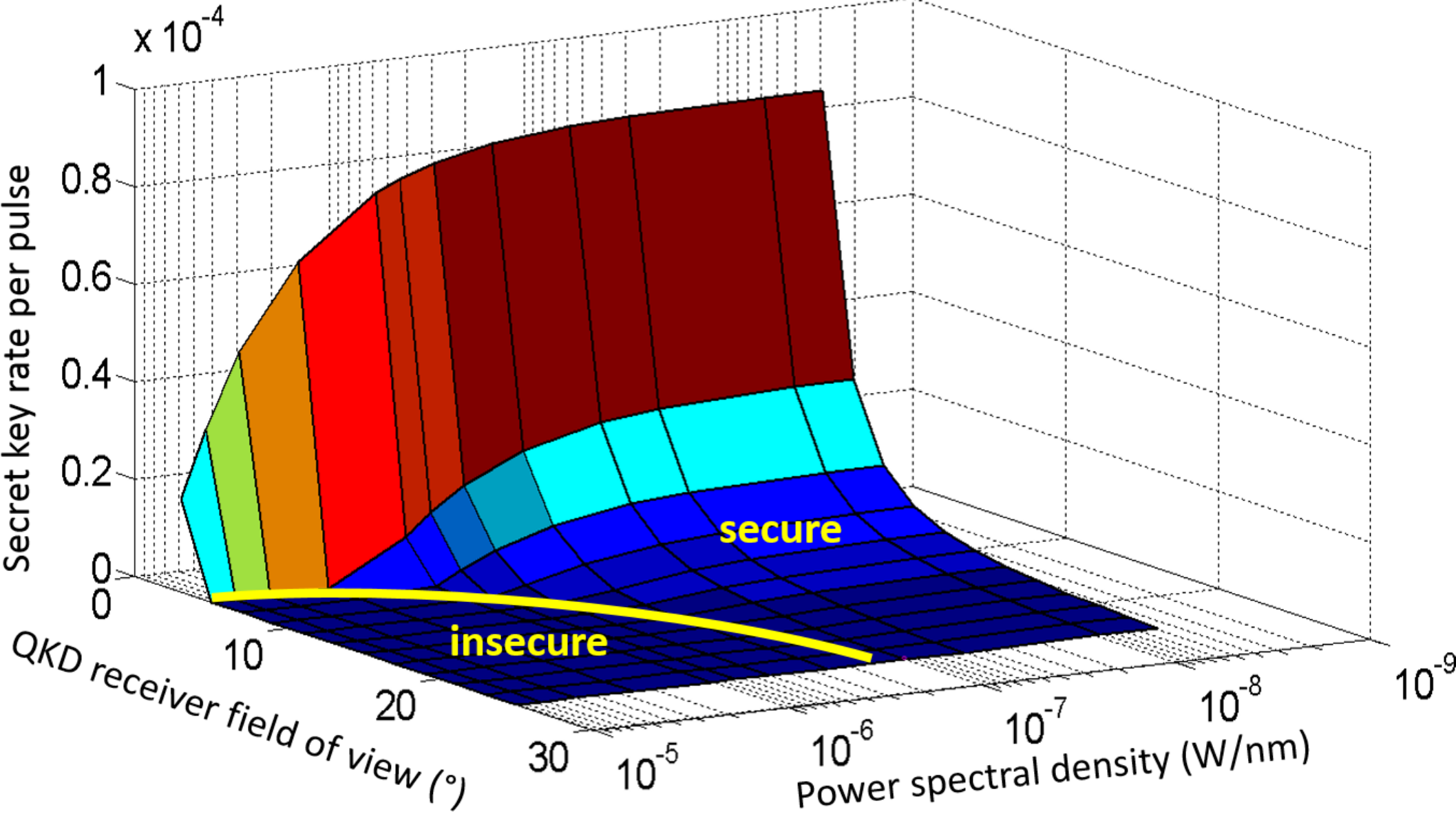}
	\caption{Secret key rate per transmitted pulse for a QKD source with ($\Theta_{1/2}=30\degree$) in a corner of the room in the presence of a lighting source. {The decoy-state BB84 protocol with an infinite number of decoy states and perfect encoders are employed.}}
	\label{fig_key_rate_3D_corner}
\end{figure}

Figures \ref{fig_key_rate_3D_center} and \ref{fig_key_rate_3D_corner} imply that while there are certain regions in which secret key exchange is possible with minimal beam alignment, additional beam steering can substantially improve the system performance. At the source, this can be achieved by narrowing the light beam and directing it toward the QKD receiver \cite{gomez2015beyond}. Here, we simulate this effect by directing a beam with $\Theta_{1/2}=5\degree$ toward the QKD receiver, while the receiver's telescope orientation is fixed facing downward. As shown in Fig.~\ref{fig_key_rate_3D_corner_BS}, the system can now tolerate a larger amount of PSD in comparison to Fig.~\ref{fig_key_rate_3D_corner}, where the QKD source is sending light in a non-direct manner with respect to the QKD receiver. Alternatively, in this case, one can use a larger FOV at the receiver, which allows us to cover a larger area for a mobile user.

\begin{figure}[!t]
	\centering
	\includegraphics[width=1\linewidth]{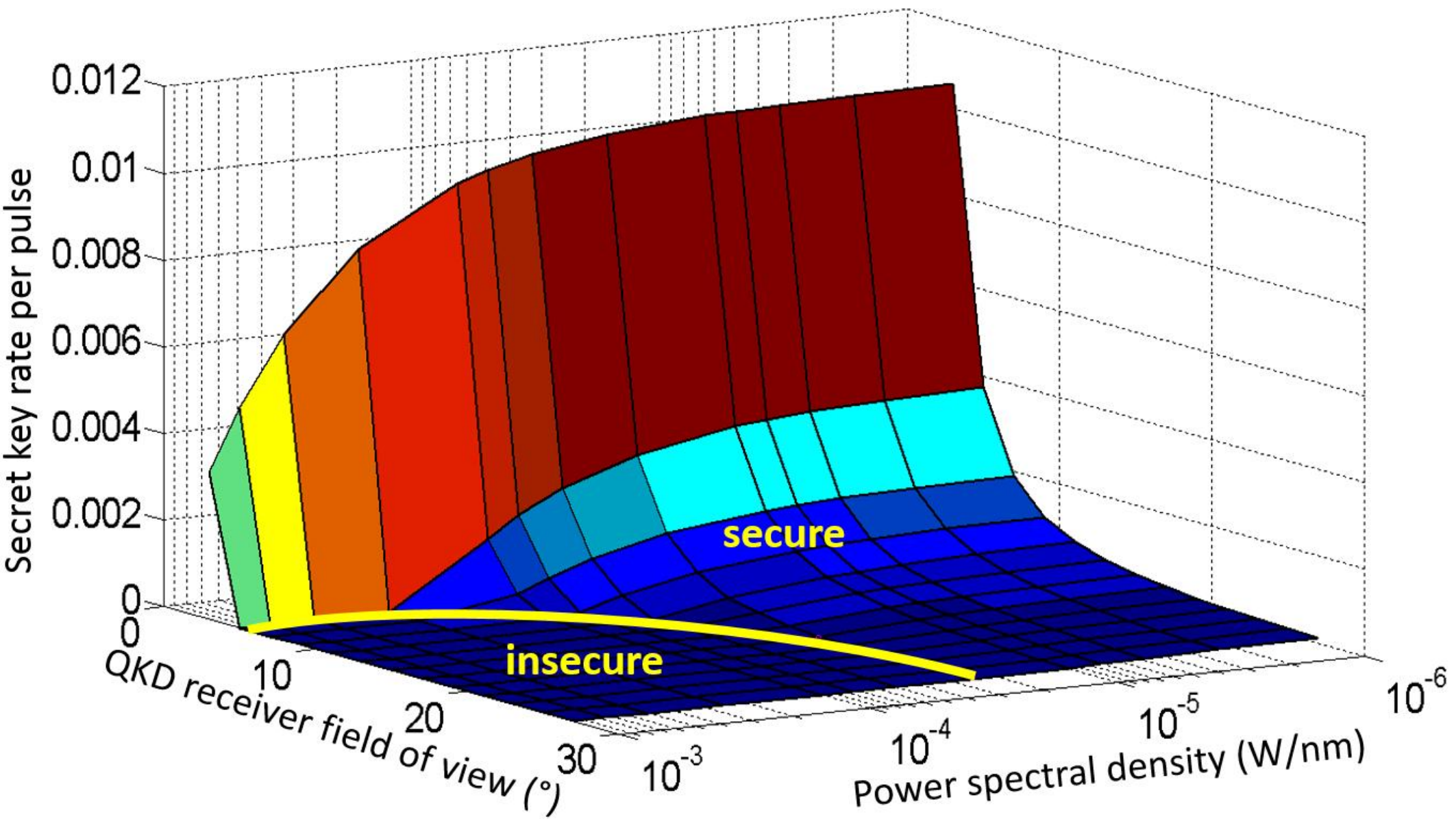}
	\caption{Secret key rate per transmitted pulse for the QKD source in a corner of the room with additional beam steering: the QKD beam is directed into the receiver with $\Theta_{1/2}=5\degree$. {The decoy-state BB84 protocol with an infinite number of decoy states and perfect encoders are employed.}}
	\label{fig_key_rate_3D_corner_BS}
\end{figure}

\subsection{More practical encoding techniques}

After assessing the feasibility of wireless indoor QKD using perfect encoders, here we study the practical cases presented in Sec.~\ref{Sec:Syst}. We consider the vacuum+weak decoy-state QKD, QKD with known source flaws, and RFI-QKD when the source flaws are not known. We use the results of Sec.~\ref{Sec:Rate} to calculate the key rate in each case. Figures~\ref{fig_practical_encoding}(a) and (b) show the secret key generation rate in each case where the QKD source is, respectively, at the center and the corner of the room. There is only loose alignment between the source and the receiver, but here we have assumed a lower PSD for the lighting source at $10^{-6}$~W/nm. In the case of known source flaws we have assumed 10\% error in the $X$ basis, whereas, in the RFI-QKD case, $X$ and $Y$ bases are rotated by a fix, but unknown, phase. In Fig.~\ref{fig_practical_encoding}(a), when the source is at the center of the room, the performance in all cases with an imperfect encoder is very close to the perfect case. This is expected as this case is less vulnerable to the alignment condition. When the source moves to the corner of the room, the sensitivity to the FOV becomes higher, but still with an FOV of $7\degree$, it is still possible to exchange secret keys. At this FOV, the drop in key rate, is around one order of magnitude when we, instead of perfect encoders, use RFI-QKD. This implies that by the proper choice of protocol we can make the system quite resilient to possible imperfections at its encoders. 

The above cases illustrate the possibility of using QKD in certain indoor environments. Whether or not we need to employ extensive beam steering in our scheme would depend on the application scenario and its target key rate. One can think of certain scenarios in which the amount of keys generated by loose beam steering would still be sufficient for the application in mind. For instance, consider a bank customer that uses an advanced encryption standard (AES) protocol, supplemented by QKD generated seeds, for his/her banking transactions. Assuming that each session roughly requires 1~kb of secret key, the user can store the amount of keys required for nearly 600 sessions within 100~s in one trip to the bank, where according to Fig.~\ref{fig_key_rate_3D_center}, a key rate of {6~kbps}, at a pulse repetition rate of 100~MHz, can be achieved when the receiver's FOV and PSD are, respectively, 5$^\circ$ and $10^{-5}$ W/nm. For applications with higher key usage, e.g., secure video streaming, one may need higher key rates only achievable if the transmitter-receiver pair are fully aligned. In such cases, one possible solution is to use docking stations, which ensures alignment without implementing all the necessary optical elements on a portable/mobile device.

\begin{figure}[!t]
	\centering
	\includegraphics[width=1\linewidth]{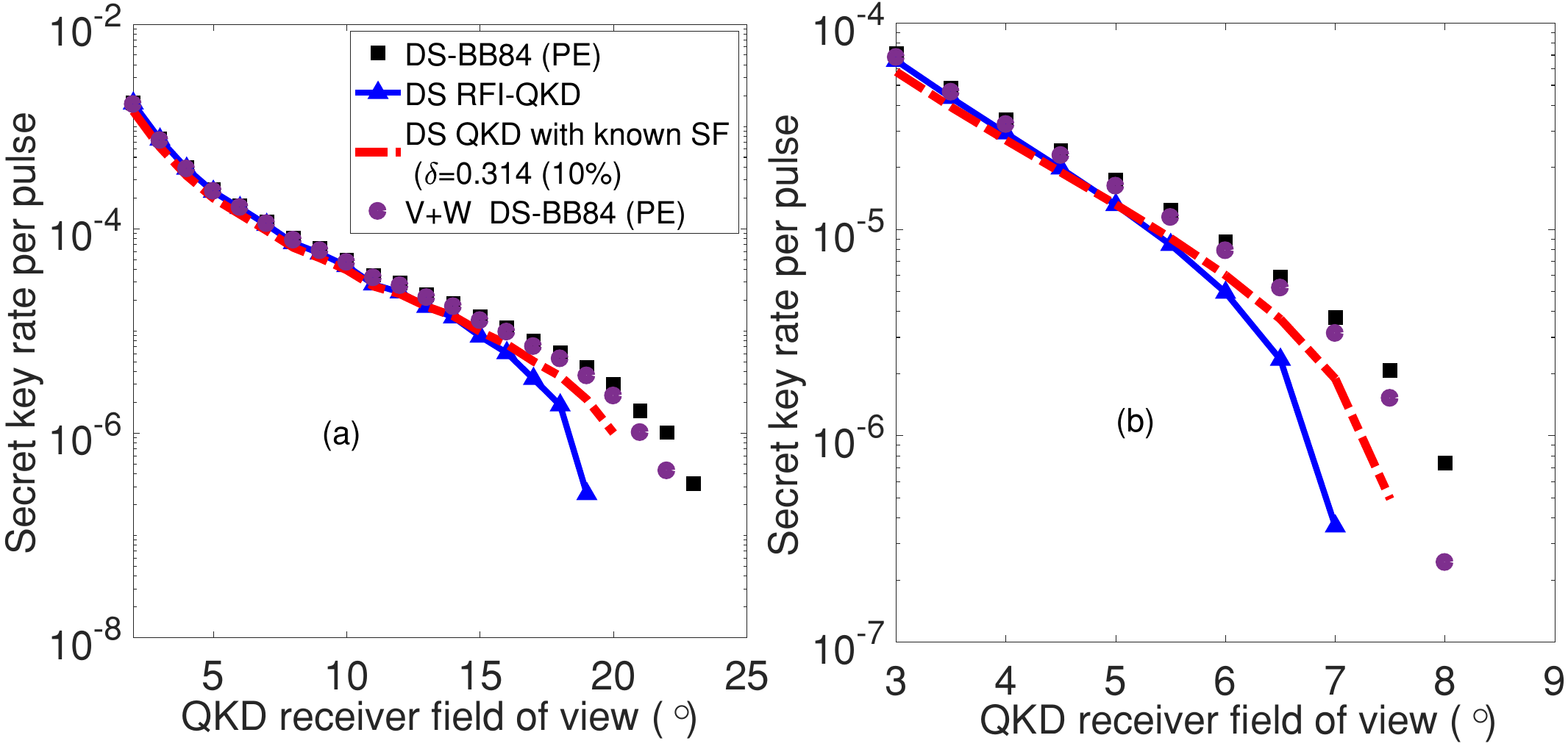}
	\caption{Secret key rate per transmitted pulse versus the QKD receiver FOV for different QKD protocols (PE: Perfect Encoding, DS: Decoy State, SF: Source Flaw, V+W: Vacuum+Weak). (a) The QKD source is located at the center of the room. (b) The QKD source is located at a corner of the room. The PSD of the bulb is $10^{-6}$~W/nm. }
	\label{fig_practical_encoding}
\end{figure}

\section{Conclusions and Discussion}
\label{Sec:Conc}
We studied the feasibility of wireless indoor QKD in a window-less room lit by an artificial source. Such systems could  provide the first link within a larger quantum network or facilitate the use of QKD in common areas for many users. We showed that there would exist a practical regime of operation within which such a wireless QKD system could generate secret keys in indoor environments. We used optical wireless communications models to characterize the path loss and background noise in the channel. Our results showed that with even mild assumptions on the alignment of the QKD transmitter and receiver, it would be possible to exchange secret keys if the room is lit by white LED bulbs. Such light sources have very little power spectral density at the operating wavelengths of interest for a QKD system, and because of their low energy consumption are expected to be ubiquitously used in the future. Our results further showed that additional enhancement could be obtained if beam steering techniques were employed. 

The type of equipment needed for the above setup is within reach of our current quantum and classical technologies. With recent progress in integrated QKD devices \cite{sibson2015chip,ma2016integrated,vest2015design}, it is possible to think of a portable device equipped with QKD capabilities. A handheld QKD prototype has, in fact, already been implemented for short-range handheld-to-ATM key exchange \cite{HP_HandheldQKD,chun2017handheld}. Because of high path loss in integrated optics systems, the integration requirement for time-bin encoding may be harsher than that of polarization encoding. One can, however, think of hybrid solutions where an integrated dual-rail setup is used for the initial encoding, which will then be converted to a single-rail time bin encoding using an external delay line. We also need random number generators~\cite{sanguinetti2014quantum}. In many scenarios, we can generate random bits offline, store them on the device, and use them during the key-exchange protocol. As for the QKD receiver, we can use some of the existing technologies for Li-Fi for collection and alignment. For instance, we can use a non-imaging optical concentrator, such as a compound parabolic collector, followed by a bandpass filter at the receiver. The output of the phase-decoding interferometer is then passed to one of the two single-photon avalanche photodiodes. Such detectors have also been considered for use in Li-Fi systems~\cite{chitnis2014spad}. Overall, with the progress made toward implementing QKD modules with integrated optics along with the progress in beam steering in classical optical communications \cite{gomez2015beyond}, our proposed system can enable high-rate wireless access to future quantum-classical networks \cite{GC2016}.


%



\section*{Acknowledgment}

The authors would like to thank Z. Ghassemlooy and D. Bitauld for fruitful discussions. All data generated in this paper can be reproduced by the provided methodology and equations.

\ifCLASSOPTIONcaptionsoff
  \newpage
\fi



%




\bibliographystyle{IEEEtran}

\bibliography{Master}

\begin{IEEEbiographynophoto}{Osama Elmabrok}
received his B.Sc degree in Electrical and Electronic Engineering from the University of Benghazi (formerly known as Garyounis University) in 2002, and M.Eng in Communication and Computer Engineering from the National University of Malaysia in 2007. He was working as an Assistant Lecturer for the University of Benghazi before joining the University of Leeds, where he is currently working toward his Ph.D. degree. Prior joining the academic field, he worked as a communications engineer for Almadar Telecom Company, Azzaawiya Refining Company, and RascomStar-QAF Company. His current research interests include quantum key distribution and its application in wireless environments.
\end{IEEEbiographynophoto}

\begin{IEEEbiographynophoto}{Mohsen Razavi}
received his B.Sc. and M.Sc. degrees (with honors) in Electrical Engineering from Sharif University of Technology, Tehran, Iran, in 1998 and 2000, respectively. From August 1999 to June 2001, he was a member of research staff at Iran Telecommunications Research Center, Tehran, Iran, working on all-optical CDMA networks and the possible employment of optical amplifiers in such systems. He joined the Research Laboratory of Electronics, at the Massachusetts Institute of Technology (MIT), in 2001 to pursue his Ph.D. degree in Electrical Engineering and Computer Science, which he completed in 2006. He continued his work at MIT as a Post-doctoral Associate during Fall 2006, before joining the Institute for Quantum Computing at the University of Waterloo as a Post-doctoral Fellow in January 2007. Since September 2009, he is a Faculty Member at the School of Electronic and Electrical Engineering at the University of Leeds. His research interests include a variety of problems in classical optical communications. In 2014, he chaired and organized the first international workshop on Quantum Communication Networks. He is currently coordinating the European Training Network QCALL, which aims at making quantum communications technologies accessible to the end users. 
\end{IEEEbiographynophoto}

\end{document}